\pdfoutput=1

\documentclass[11pt,twoside,a4paper,cmspaper,final,collab]{cms-tdr}
\def\svnVersion{59fbca1-D}\def\svnDate{2019/08/07}\def\cmsCernNoTag{CERN-EP-2019-088}\def\cmsCernDate{\today}\def\cmsMessage{"Published in Physical Review D as \href{http://dx.doi.org/10.1103/PhysRevD.100.052003}{\doi{10.1103/PhysRevD.100.052003}.}"}

\begin{document}\cmsNoteHeader{EXO-18-005}

\hyphenation{had-ron-i-za-tion}
\hyphenation{cal-or-i-me-ter}
\hyphenation{de-vices}
\newcommand{\fullLumi}{77.4\fbinv}
\newcommand{\ptvecl}{\ensuremath{\ptvec^{\kern1pt\ell}}\xspace}
\newcommand {\dr}{\ensuremath{\Delta R}\xspace}
\newcommand{\WZ}{\ensuremath{\PW\PZ}\xspace}
\newcommand{\ZZ}{\ensuremath{\PZ\PZ}\xspace}
\newcommand{\LT}{\ensuremath{L_\mathrm{T}}\xspace}
\newcommand{\cPV}{\HepParticle{V}{}{}{}\xspace}
\newcommand{\cPX}{\HepParticle{X}{}{}{}\xspace}
\newcommand\hairspace{\kern .08333em }

\newlength\cmsFigWidth
\newlength\cmsTabSkip\setlength{\cmsTabSkip}{1ex}
\ifthenelse{\boolean{cms@external}}{\setlength\cmsFigWidth{0.85\columnwidth}}{\setlength\cmsFigWidth{0.4\textwidth}}
\ifthenelse{\boolean{cms@external}}{\providecommand{\cmsLeft}{top\xspace}}{\providecommand{\cmsLeft}{left\xspace}}
\ifthenelse{\boolean{cms@external}}{\providecommand{\cmsRight}{bottom\xspace}}{\providecommand{\cmsRight}{right\xspace}}
\ifthenelse{\boolean{cms@external}}{\providecommand{\CL}{C.L.\xspace}}{\providecommand{\CL}{CL\xspace}}
\providecommand{\cmsTable}[1]{\resizebox{\textwidth}{!}{#1}}
\cmsNoteHeader{EXO-18-005}
\title{Search for vector-like leptons in multilepton final states in proton-proton collisions at \texorpdfstring{$\sqrt{s} = 13$\TeV}{sqrt(s) = 13 TeV}}

\author*[inst1]{Anshul Kapoor}

\date{\today}

\abstract{A search for vector-like leptons in multilepton final states is presented. The data sample corresponds to an integrated luminosity of \fullLumi  of proton-proton collisions at a center-of-mass energy of 13\TeV collected by the CMS experiment at the LHC in 2016 and 2017. Events are categorized by the multiplicity of electrons, muons, and hadronically decaying \Pgt leptons. The missing transverse momentum and the scalar sum of the lepton transverse momenta are used to distinguish the signal from background. The observed results are consistent with the expectations from the standard model hypothesis. The existence of a vector-like lepton doublet, coupling to the third generation standard model leptons in the mass range of 120--790\GeV, is excluded at 95\% confidence level. These are the most stringent limits yet on the production of a vector-like lepton doublet, coupling to the third generation standard model leptons.
}

\hypersetup{%
pdfauthor={CMS Collaboration},%
pdftitle={Search for vector-like leptons in multilepton final states in pp collisions at sqrt{s} = 13 TeV},%
pdfsubject={CMS},%
pdfkeywords={CMS, vector-like leptons}}

\maketitle

\section{Introduction}\label{sec:intro}

The standard model (SM) of particle physics is a quantum field theory that describes the known fundamental particles and
their interactions. The predictions of the SM have been experimentally tested with great precision~\cite{Tanabashi:2018oca}.
However, the SM does not explain several observations, such as the existence of dark matter and the baryon asymmetry in the
universe. In addition, there exist theoretical issues such as the hierarchy problem, that suggest that an extension of the SM, predicting new particles, is needed to provide a more complete description of nature.

In one class of new particles there are nonchiral color singlet fermions that couple to the SM leptons. The term nonchiral implies that
the left- and right-handed components of these particles transform identically under gauge symmetries. These particles are thus referred
to as vector-like leptons (VLLs). They arise in a wide variety of models invoking, for example, supersymmetry or extra dimensions~\cite{Martin:2009bg, Endo:2011mc, Dermisek:2013gta, Halverson:2014nwa}. The VLLs are often classified by the SM lepton generation with which they are associated. VLLs and their associated SM leptons have identical lepton numbers.

\begin{figure*}[!ht]
\centering
\includegraphics[width=0.4\textwidth]{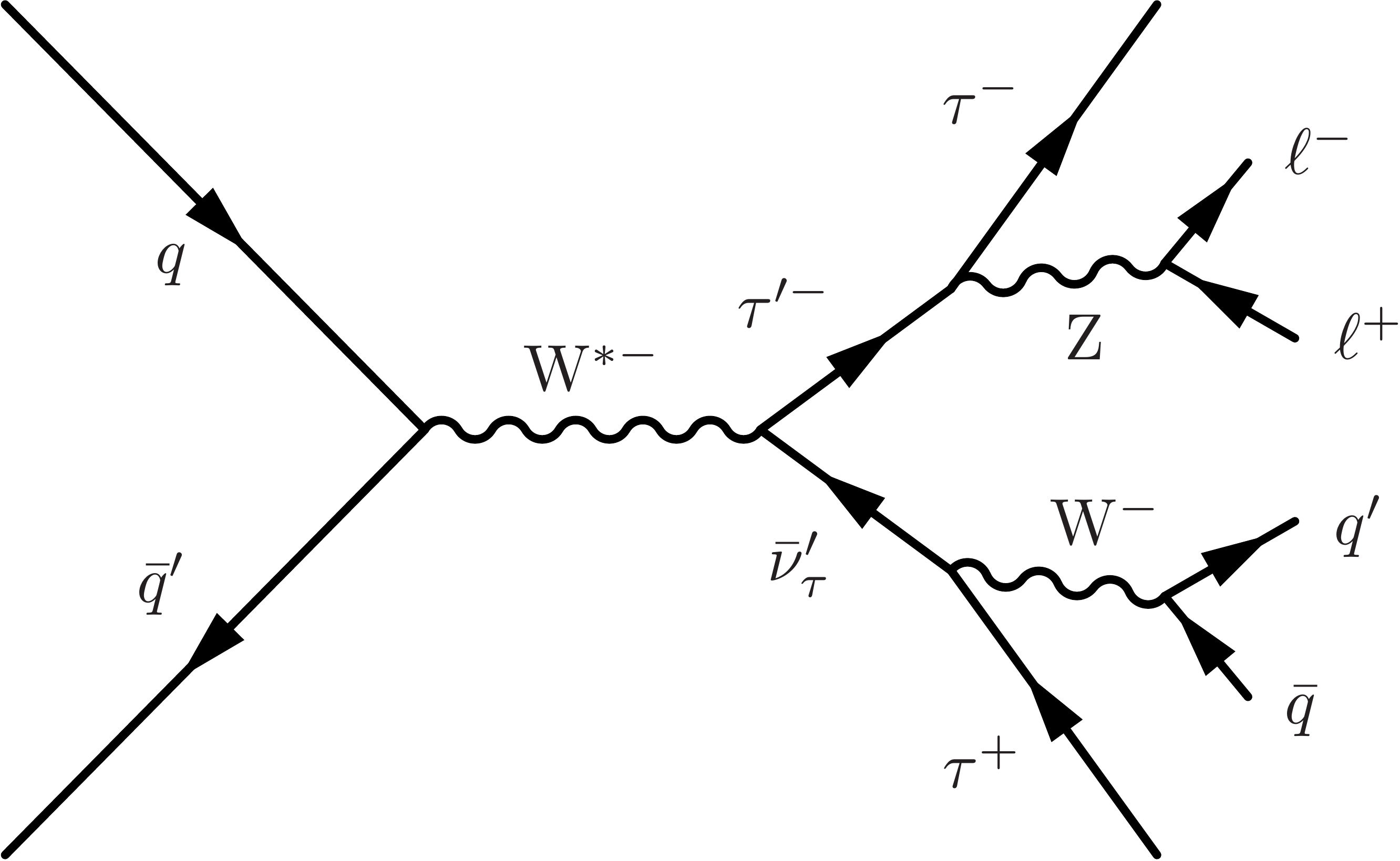}
\includegraphics[width=0.4\textwidth]{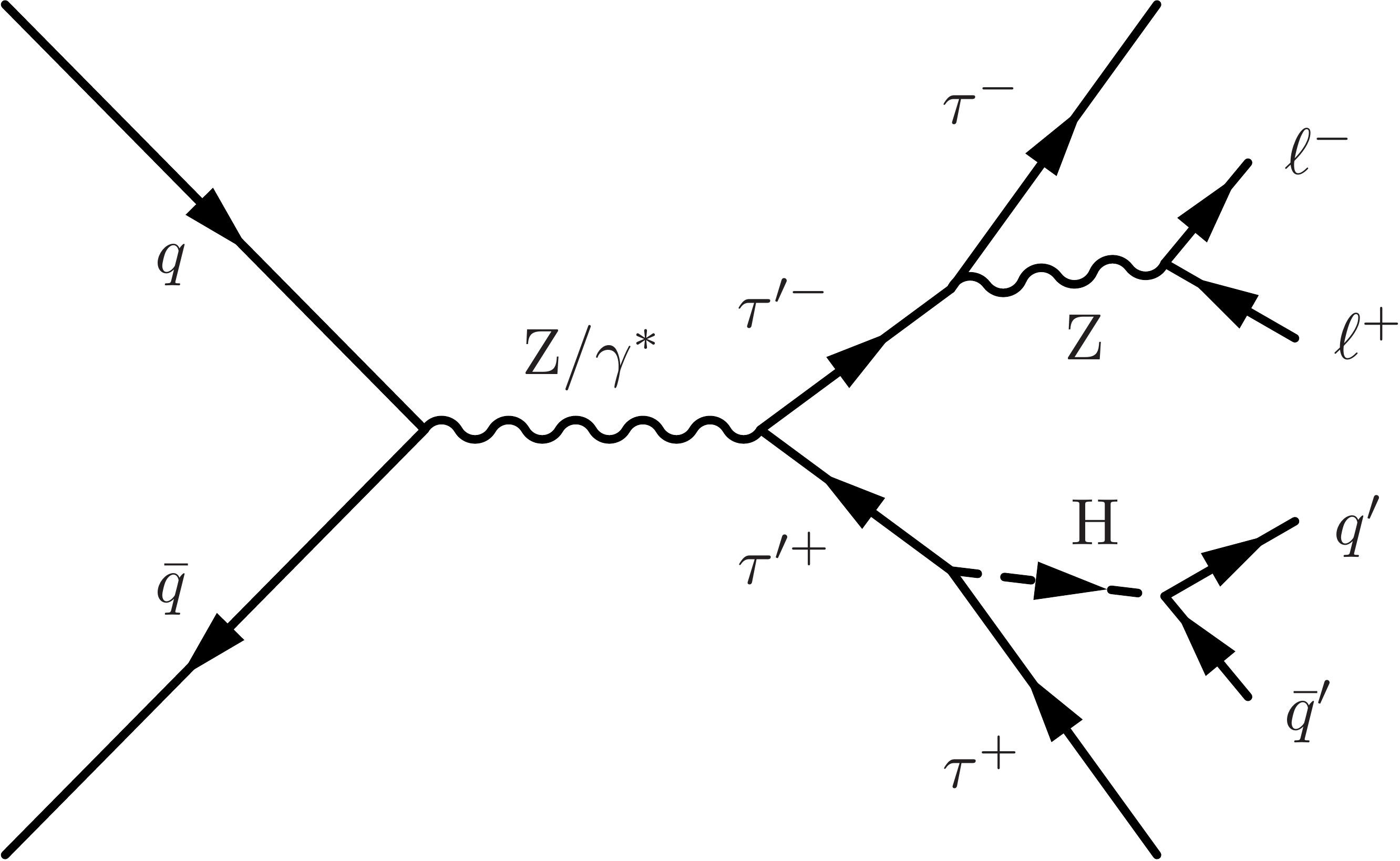}
\caption{Two illustrative leading order Feynman diagrams for associated production of $\Pgt '$ with a $\PGn^\prime_{\Pgt}$ (left) and for pair production of $\Pgt '$ (right), and possible subsequent decay chains that result in a multilepton final state.}
\label{fig:VLLFeyn}
\end{figure*}

This paper presents a search for an SU(2) doublet VLL extension~\cite{Kumar:2015tna} of the SM with couplings to the third
generation SM leptons. The search is carried out in final states with multiple charged leptons (\Pe, \Pgm, \Pgt), using proton-proton ($\Pp\Pp$) collision data collected by the CMS detector at the LHC in 2016 and 2017. The model that we consider introduces a
vector-like \Pgt lepton $(\Pgt '^{-})$, its antiparticle $(\Pgt '^{+})$, and the corresponding neutrinos ($\PGn^\prime_{\Pgt}$ and $\PAGn^\prime_{\Pgt}$). At the LHC, they can be produced in $\Pgt '^{\pm}\PGn^\prime_{\Pgt}$, $\Pgt '^{+}\Pgt '^{-}$, and $\PGn^\prime_{\Pgt}\PAGn^\prime_{\Pgt}$ channels, with subsequent decays of $\Pgt '$ to \PZ\Pgt or \PH\Pgt and of $\PGn^\prime_{\Pgt}$ to \PW\Pgt, where \PW, \PZ, and \PH are the SM \PW, \PZ, and Higgs bosons. At tree-level, the $\Pgt '$ and  $\PGn^\prime_{\Pgt}$ are mass degenerate, whereas higher order radiative corrections predict $<$0.3\% relative mass splitting between these two states, for VLL masses greater than 100\GeV. In this paper, $\Pgt '$ and  $\PGn^\prime_{\Pgt}$ are assumed to be mass degenerate. The mass of the VLL is the only free parameter both in the production cross section and in the branching fraction calculations.  The tree-level Feynman diagrams for associated and pair production of the doublet model VLLs are shown in Fig.~\ref{fig:VLLFeyn} along with possible subsequent decay chains that would result in a multilepton final state.

The ATLAS Collaboration performed a search for heavy lepton resonances decaying into a \PZ boson and a lepton in a multilepton final state at a center-of-mass energy of 8\TeV~\cite{Aad:2015dha}, constraining a singlet VLL model and excluding VLLs in the mass range of 114--176\GeV. However, to date, there are no such constraints on the doublet VLL model from any of the LHC experiments. The L3 Collaboration at LEP placed a lower bound of $\approx$100\GeV on additional heavy leptons~\cite{Achard:2001qw}. Given these existing constraints, this analysis focuses on VLL masses greater than 100\GeV.

\section{The CMS detector}
\label{sec:detector}

The central feature of the CMS apparatus is a superconducting solenoid of 6\unit{m} internal diameter, providing a magnetic field of 3.8\unit{T}. Within the solenoid volume are a silicon pixel and strip tracker, a lead tungstate crystal electromagnetic calorimeter (ECAL), and a brass and scintillator hadron calorimeter, each composed of a barrel and two endcap sections. Muons are measured in gas-ionization detectors embedded in the steel flux-return yoke outside the solenoid. The inner tracker measures charged particles with $\abs{\eta}<2.5$ and provides an impact parameter resolution of $\approx$15\mum and a transverse momentum (\pt) resolution of about 1.5\% for 100\GeV charged particles. Extensive forward calorimetry complements the barrel and endcap detectors by covering the pseudorapidity range $3.0 < \abs{\eta} <5.2$. Collision events of interest are selected using a two-tiered trigger system~\cite{Khachatryan:2016bia}. The first level, composed of custom hardware processors, selects events at a rate of around 100\unit{kHz}. The second level, based on an array of microprocessors running a version of the full event reconstruction software optimized for fast processing, reduces the event rate to around 1\unit{kHz} before data storage. A detailed description of the CMS detector, along with a definition of the coordinate system and relevant kinematic variables, can be found in Ref.~\cite{Chatrchyan:2008aa}.

\section{Event reconstruction and particle identification}
\label{sec:obj}

Events collected for this search are recorded using a combination of triggers requiring a single electron or a single muon. For events collected in 2016 (2017), the electron trigger requires an electron with $\pt > 27$ (35)\GeV, while the muon trigger requires a muon with $\pt > 24$ (27)\GeV. Information from all subdetectors is combined using the CMS particle-flow (PF) algorithm~\cite{Sirunyan:2017ulk} to reconstruct and identify individual particles (charged hadrons, neutral hadrons, photons, electrons, and muons). Collectively these are referred to as PF objects.

For each event, PF objects originating from the same interaction vertex are clustered into jets using the infrared- and collinear-safe anti-\kt algorithm~\cite{Cacciari:2008gp,Cacciari:2011ma}, with a radius parameter of 0.4. The momenta of all PF objects in each jet are summed vectorially to determine the jet momentum. The reconstructed vertex with the largest value of summed physics-object $\pt^2$ is taken to be the primary $\Pp\Pp$ interaction vertex. The physics objects are the jets, clustered using the jet finding algorithm~\cite{Cacciari:2008gp,Cacciari:2011ma} with the tracks assigned to the vertex as inputs, and the associated missing transverse momentum, taken as the negative vector sum of the \pt of those jets. Additional interactions within the same or nearby bunch crossings (pileup) can contribute spurious extra tracks and calorimetric energy depositions to the jet momentum. Hence, charged particles identified as originating from pileup vertices are discarded and an offset correction~\cite{Cacciari:2007fd} is applied to account for the remaining neutral pileup particle contributions. Additional jet energy corrections are applied to account for the nonlinear response of the detectors~\cite{CMS-DP-2018-028}.

The missing transverse momentum vector (\ptvecmiss) is calculated as the negative vectorial \pt sum of all the PF objects belonging to the primary vertex. The \ptmiss is defined as the magnitude of this vector. For calculating \ptmiss in 2016, we use PF objects located in the full fiducial
volume of the detector, whereas for 2017, PF objects within $2.5 < \abs{\eta} < 3.0$ and with $\pt < 50 \GeV$ are excluded to mitigate noise effects
related to the aging of the CMS ECAL.

Electron candidates are reconstructed by combining ECAL superclusters and Gaussian sum filter~\cite{Adam_2005} tracks from the silicon
tracker~\cite{Khachatryan:2015hwa}. Muon candidates are reconstructed by combining the information from both the silicon tracker and the
muon spectrometer~\cite{Sirunyan:2018fpa}. Hadronically decaying \Pgt lepton candidates (\tauh) are selected using the
hadron-plus-strips algorithm~\cite{Khachatryan:2015dfa}.
This algorithm has been designed to optimize the performance of \tauh reconstruction by considering specific \tauh decay modes.
It starts with hadronic jets and reconstructs \tauh candidates from the tracks (``prongs'') and energy deposits in strips of the
ECAL, in the 1-prong, 1-prong + $\pi^{0}$, and 3-prong decay modes. We require the reconstructed leptons to lie within the region of pseudorapidity $\abs{\eta}< 2.5$, $2.4$, and $2.3$ for the electron, muon, and \tauh candidates, respectively.

Lepton candidates arising from $\Pp\Pp$ collisions can be broadly categorized into prompt, nonprompt, and conversion leptons.
A prompt lepton can be produced in the decay of a \PW, \PZ or Higgs boson. Events from background processes such as \WZ and \ZZ contain multiple
prompt leptons and thus these backgrounds are classified as prompt backgrounds. A nonprompt lepton can arise in heavy flavor hadron decays within a jet, or from hadrons that punch through to the muon system, or from hadronic showers with large electromagnetic fractions. A small fraction
of reconstructed leptons from nonprompt sources mimic leptons from prompt sources and are referred to as misidentified leptons. The background arising
from such sources is referred to as the misidentified background (MisID). A conversion lepton is one which is produced when a radiated photon
converts to a pair of leptons. The background arising from such processes is referred to as the conversion background.

Unlike prompt leptons, misidentified leptons are expected to have significant nearby hadronic activity. An isolation requirement
that compares the \pt of a lepton to the \pt sum of particles in its immediate neighborhood strongly reduces the backgrounds
from misidentified leptons. We use relative isolation criteria for both electrons and muons. Relative isolation is defined as the
scalar \pt sum of photons, and charged and neutral hadrons, as reconstructed by the PF algorithm within a specified \dr cone around
the lepton candidate, normalized to the lepton candidate $\pt$. The \dr between a particle and the lepton is defined as
$\dr = \sqrt{\smash[b]{(\Delta \eta)^2 + (\Delta \phi)^2}}$, where $\Delta \eta$ is the difference in pseudorapidity, and $\Delta \phi$ is
the difference in the azimuthal angle (in radians). This relative isolation is required to be less than 7 or 8\% within a cone of size $\dr=0.3$ for electrons whose energy deposits are reconstructed in the ECAL barrel ($\abs{\eta}<1.48$) or in the endcap ($1.48<\abs{\eta}<3.00$), respectively, and less than 15\% within a cone of size $\dr=0.4$ for muons. The \tauh candidates are required to pass an isolation requirement based on a multivariate analysis~\cite{Sirunyan:2018pgf}. The isolation quantities are corrected for pileup by considering only those charged PF candidates that are consistent with having originated from the primary vertex, and by subtracting a per-event average pileup contribution to the neutral PF components. We further reduce the MisID backgrounds by imposing requirements on the longitudinal ($d_z$) and transverse ($d_{xy}$) impact parameters of the leptons with respect to the primary vertex in the event. Electrons in the barrel (endcap) must satisfy $\abs{d_z} < 0.1$ $(0.2)\cm$ and $\abs{d_{xy}} <0.05$ $(0.1)\cm$. Muons must satisfy $\abs{d_{z}} < 0.1\cm$ and $\abs{d_{xy}} < 0.05\cm$. For \tauh leptons, we require $\abs{d_z} < 0.2\cm$.

\section{Signal and background simulation}\label{sec:backgrounds}

Simulated samples are used to estimate the contribution of all prompt and conversion background processes. The \WZ and \ZZ processes
are generated at next-to-leading order (NLO)  using \POWHEG v2~\cite{Nason:2004rx, Frixione:2007vw, Alioli:2010xd, Melia2011, Nason:2014}. The $\PZ/\gamma^*$, $\PZ/\gamma^* + \gamma$, \ttbar, $\ttbar + \gamma$, and triboson processes are generated at NLO using \MGvATNLO v5.2.2~\cite{Alwall:2014hca} and processes with the Higgs boson are generated using \POWHEG v2~\cite{Bagnaschi2012, Klasen2012} and the \textsc{JHUGen} v6.2.8 generator~\cite{Gao:2010qx, Bolognesi:2012mm, Anderson:2013afp, Gritsan:2016hjl}. Signal events are generated using \MGvATNLO at leading order (LO) precision. For all simulation data, the parton showering, fragmentation, and hadronization steps are done using \PYTHIA 8.230~\cite{Sjostrand:2014zea} with tune CUETP8M1~\cite{Khachatryan:2015pea} for 2016 samples, and CP5~\cite{CMS:2018zub} for 2017 samples.

All 2016 samples are generated with the same order of the NNPDF3.0 parton distribution function (PDF)~\cite{Ball:2014uwa} as the order of the MC generator. All 2017 samples are generated with the NNPDF3.1 next-to-next-to-leading (NNLO) order PDF~\cite{Ball:2017nwa}, irrespective of the order of the MC generator. The response of the CMS detector is simulated using dedicated software based on the \GEANTfour toolkit~\cite{Agostinelli:2002hh}. Additional weights are applied to all simulated events to account for differences in the trigger and lepton identification efficiencies between data and simulation. For the simulated events, additional minimum bias interactions are superimposed on the primary collision, reweighted in such a way that the frequency distribution of the extra interactions matches that observed in data.

\section{Event selection criteria}\label{sec:eventSelections}

We collectively refer to electrons and muons as light-leptons to distinguish them from \tauh leptons. Events are then categorized as those with four or more light-leptons (4L), exactly three light-leptons (3L), and exactly two light-leptons along with at least one \tauh lepton (2L1T). In the 2L1T channel, we have a further division based on whether the two light-leptons are of opposite sign (OS) or same sign (SS). In all categories, the leptons are ordered by decreasing transverse momenta and those with the largest \pt are labeled as the leading leptons. The leading light-lepton is required to satisfy $\pt > 38$ (28)\GeV if it is an electron (muon). These thresholds are imposed so that the corresponding single lepton triggers are fully efficient for events that would subsequently satisfy the offline selection. All of the other leptons are required to satisfy $\pt > 20 \GeV$.

We use the scalar \pt sum of the leptons (denoted as \LT) to discriminate signal from SM backgrounds in all channels. The
\LT distribution is divided into 150\GeV bins, each of which is treated as a separate experiment. In the
2L1T and 4L categories that contain more than one \tauh and more than four light-lepton candidates, respectively, only the
leading \tauh and the leading four light-leptons are used in the calculation of \LT.

In order to improve sensitivity for the signal, in each of the 4L, 3L, and 2L1T (OS, SS) categories, the events are divided
into low- and high-\ptmiss regions. While the 4L category is divided into $\ptmiss<50 \GeV$ and $>$50\GeV regions, the 3L and 2L1T (OS, SS)
categories are divided into $\ptmiss<150\GeV$ and $>$150\GeV regions. These categories form the bases of signal regions (SR) that would be sensitive to the presence
of a VLL signal. They are complemented by orthogonal control regions (CR) that are expected to be dominantly populated by backgrounds.
Additionally, all events with a light-lepton pair invariant mass below 12\GeV are vetoed regardless of the flavor and sign of the pair,
in order to suppress low mass quarkonia resonances. The SRs are described in Table~\ref{tab:SR}, where OSSF refers to an opposite-sign,
same-flavor lepton pair. A detailed description of the CRs is given in Section~\ref{sec:backgroundEstimation}.

\begin{table*}[!ht]
\centering
\topcaption{The signal regions defined in this analysis. The on-\PZ mass window is defined as $76< m_{\ell\ell} < 106 \GeV$, while the below-\PZ condition
  is defined as $m_{\ell\ell}< 76 \GeV$.}
\begin{scotch}{ccc}
  N$_\text{leptons}$ & \ptmiss ({\GeVns}) & CR veto \\
  \hline

\multirow{2}{*}{$\geq$4$\Pe / \Pgm$} & $<$50 & \multirow{2}{*}{2 OSSF on-\PZ pairs and $\ptmiss<50 \GeV$} \\
& $>$50 & \\[\cmsTabSkip]

\multirow{3}{*}{$3 \Pe / \Pgm$} & \multirow{2}{*}{$<$150} & OSSF on-\PZ pair and $\ptmiss < 100 \GeV$, or \\

& \multirow{2}{*}{$>$150}& OSSF below-\PZ pair and $\ptmiss<50 \GeV$, or \\
&  & OSSF below-\PZ pair and on-\PZ $m_{3\ell}$\\[\cmsTabSkip]

\multirow{2}{*}{$2 \Pe/\Pgm$ OS (or SS) + $\geq$1\tauh} & $<$150 & \multirow{2}{*}{$\ptmiss<50 \GeV$} \\
& $>$150 & \\
\end{scotch}
\label{tab:SR}
\end{table*}

\section{Background estimation}\label{sec:backgroundEstimation}

The \WZ and \ZZ background yields are normalized to data using dedicated CRs. For the \WZ CR, we select events
with exactly three light-leptons, one OSSF pair invariant mass satisfying the $91\pm15 \GeV$ window (``on-\PZ''), and $50 < \ptmiss < 100 \GeV$.
The ratio of the expected \WZ yield to data (after correcting for non-\WZ events) is found to be $1.14\pm 0.06$ ($1.07\pm 0.05$) for the 2016 (2017)
data analysis, where the uncertainty includes both statistical and systematic contributions.
Similarly, for the \ZZ background, we select events with exactly four leptons, two distinct OSSF pairs both satisfying the on-\PZ requirement,
and $\ptmiss < 50 \GeV$. The ratio of the expected \ZZ yield to data is found to be $1.01\pm 0.05$ ($0.98\pm 0.05$) for the 2016 (2017) search.

The conversion background consists of events with photons from final-state radiation, where the photon converts asymmetrically to two
additional leptons, only one of which is reconstructed in the detector. A selection of events with three light-leptons with an OSSF pair below the \PZ boson mass ($<$76\GeV), M$_{3\ell}$ satisfying on-\PZ window, and with $\ptmiss<50 \GeV$ is used to calculate the ratio of the conversion background prediction in simulation to data. The quantity $m_{3\ell}$ is defined as the invariant mass of the three light-leptons. The ratio is measured to be $0.95\pm 0.11$ ($0.87\pm 0.10$) for the 2016 (2017) data analysis. For the 2017 analysis, the $\PZ/\gamma^* + \gamma$ and  $\ttbar+\gamma$ simulation samples are used, while for the 2016 analysis, the $\PZ/\gamma^*$ and $\ttbar$ simulation samples are used because of the unavailability of enhanced samples.

\begin{figure*}[!ht]
  \centering
  \includegraphics[width=0.49\textwidth]{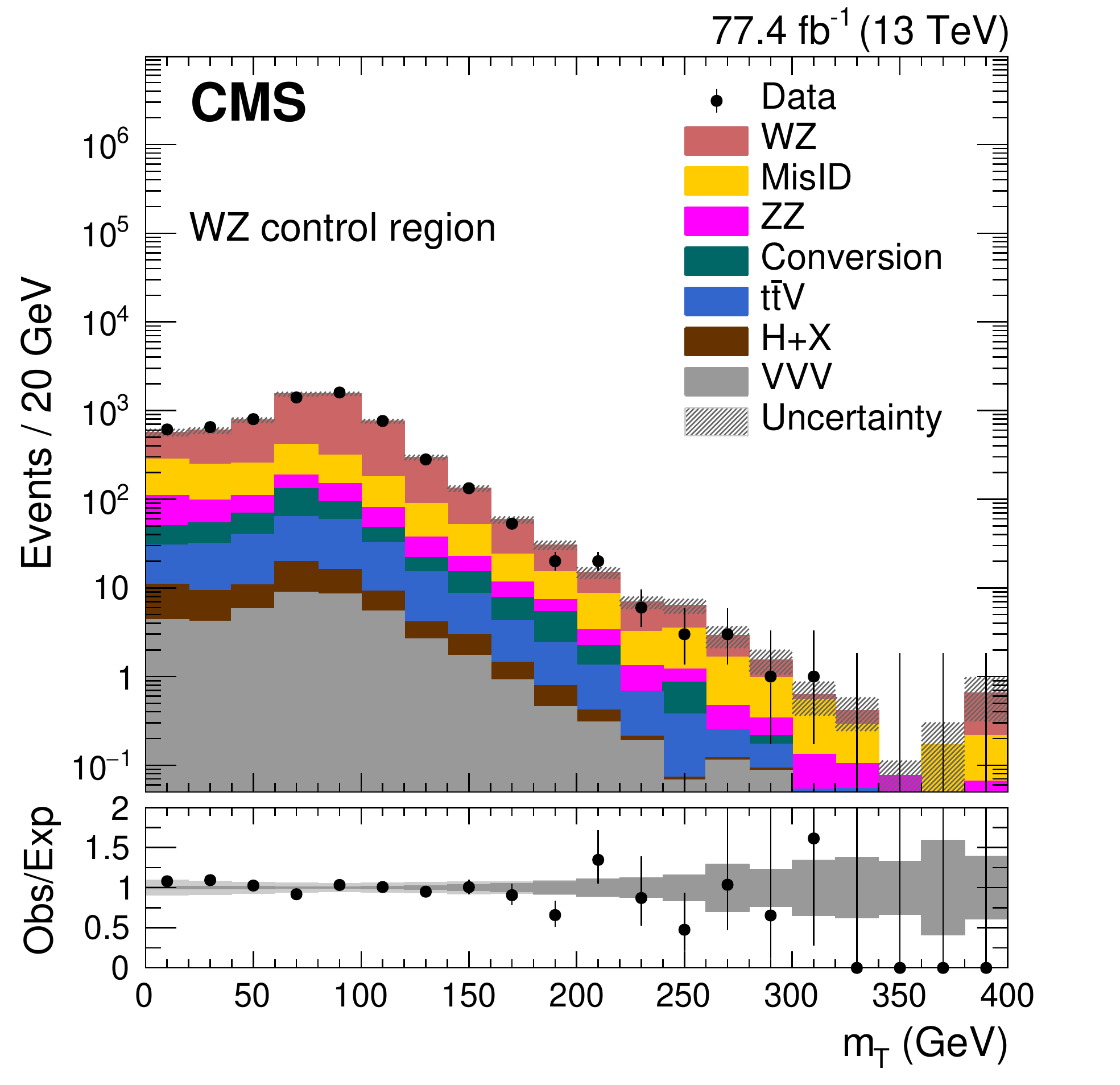}
  \includegraphics[width=0.49\textwidth]{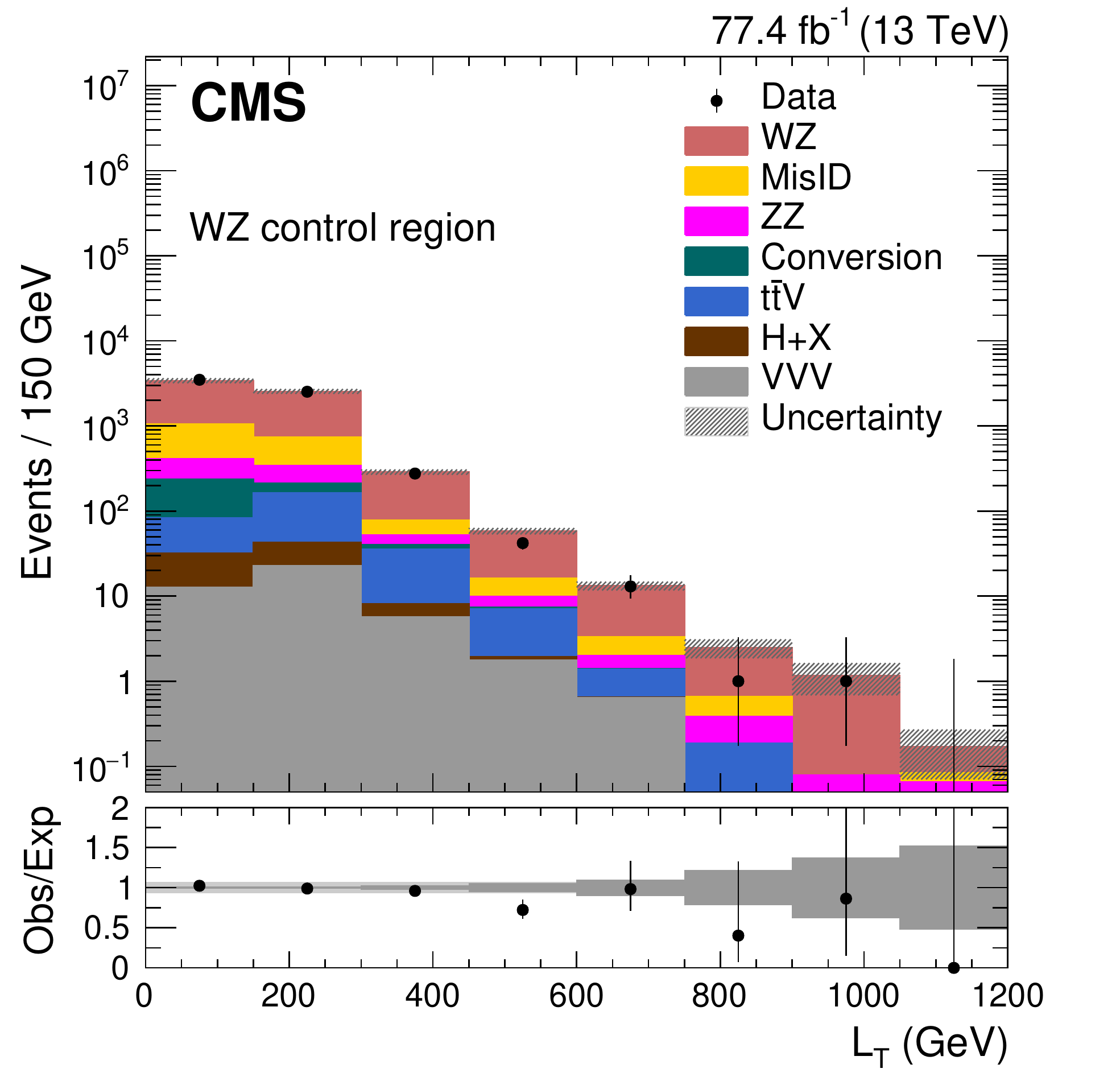}
  \includegraphics[width=0.49\textwidth]{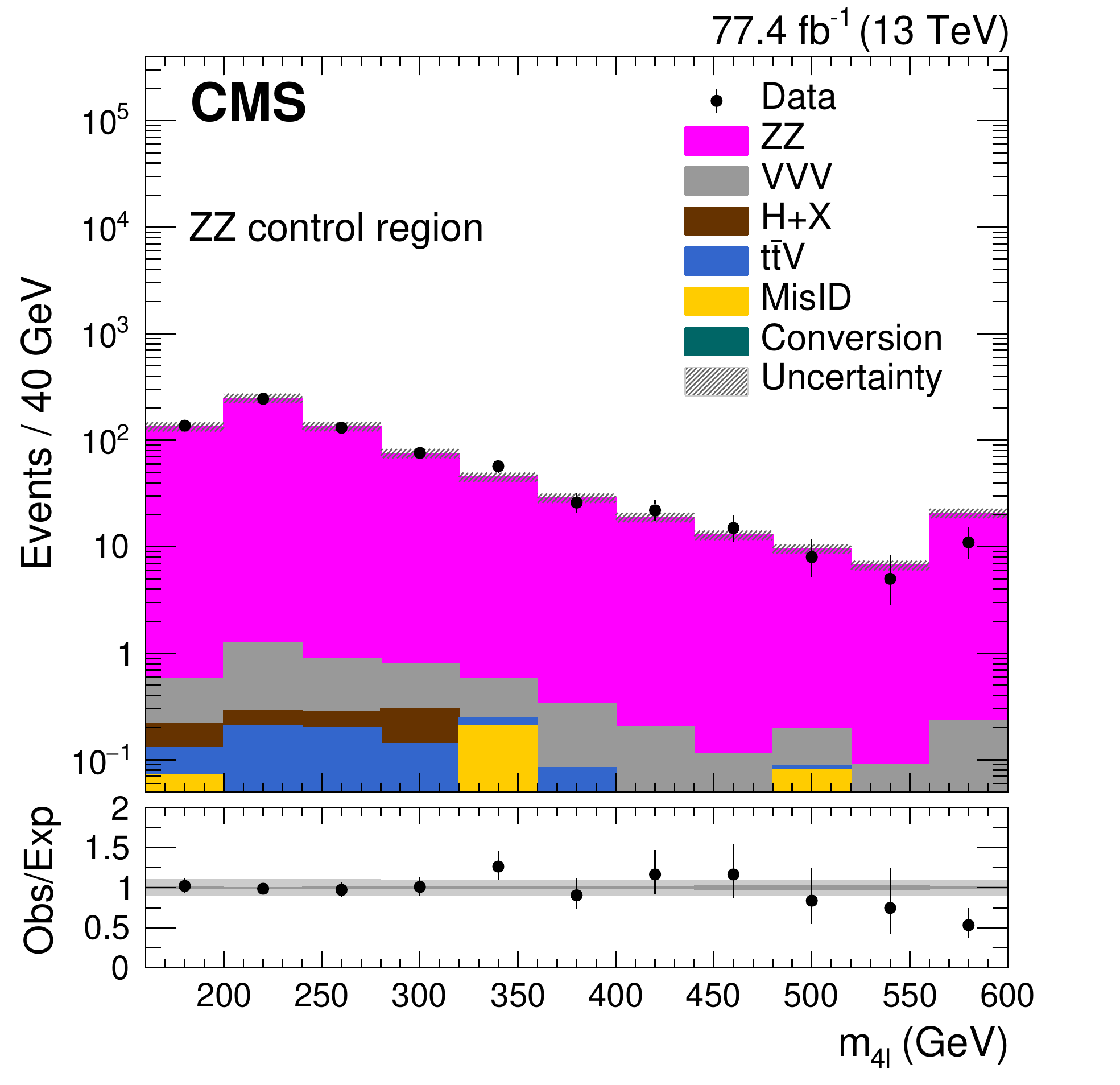}
  \includegraphics[width=0.49\textwidth]{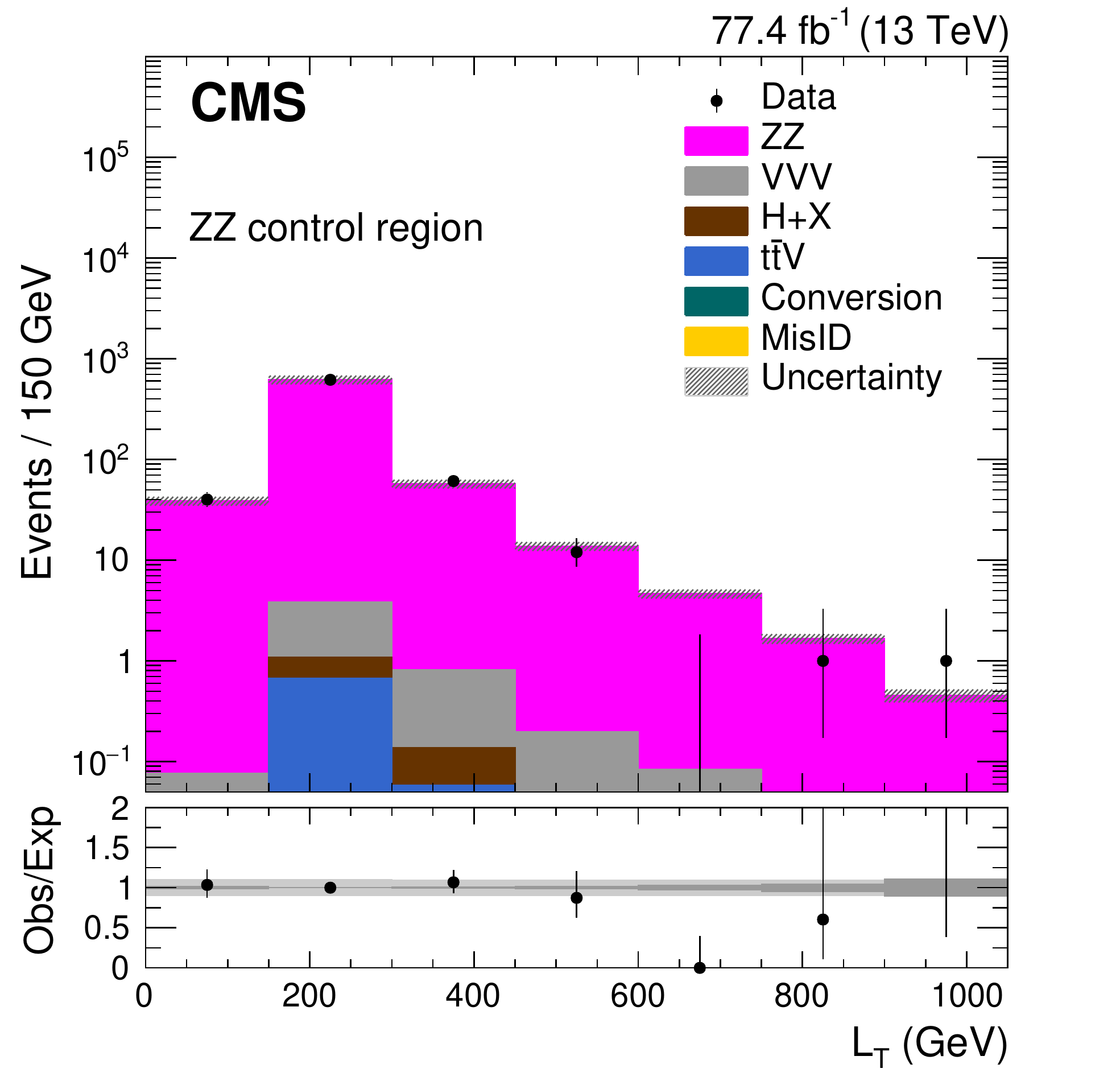}
  \caption{\label{fig:wzzz} The upper row shows the \mT (left) and the \LT (right) distributions in the \WZ control region in data and simulation. The \WZ control region contains events with three leptons and an OSSF pair with mass on-\PZ, and $50< \ptmiss < 100 \GeV$. The lower row shows the $m_{4\ell}$ (left) and the \LT (right) distributions in the \ZZ control region. The \ZZ control region contains events with two OSSF lepton pairs, both of which are on-\PZ, and $\ptmiss < 50 \GeV$. The total SM background is shown as a stack of all contributing processes. The hatched gray bands in the upper panels represent the total uncertainty in the expected background. The lower panels show the ratios of observed data to the total expected background. In the lower panels, the light gray band represents the combined statistical and systematic uncertainty in the expected background, while the dark gray band represents the statistical uncertainty only. The rightmost bins include the overflow events.}
\end{figure*}

The measured ratios are then applied to the \WZ, \ZZ, and conversion background estimates to correct for any residual differences in
the efficiency and acceptance between data and simulation. The CRs are also used to verify the performance of the simulation in modeling the kinematic distributions of interest. Figure~\ref{fig:wzzz} shows the transverse mass \mT and the \LT distributions in the \WZ CR and the $m_{4\ell}$ and \LT distributions in the \ZZ CR for data and simulation, in the combined 2016 and 2017 data sets. The quantity $m_{4\ell}$ is defined as the invariant mass of the leading four light-leptons. The quantity \mT is defined as $\mT = \sqrt{\smash[b]{2 \ptmiss \pt^{\,\ell} [1 - \cos(\Delta\phi_{\mT})]}}$ where $\pt^{\,\ell}$ refers to the \pt of the lepton that is not part of the OSSF pair closest to the \PZ boson mass and $\Delta\phi_{\mT}$ is the difference in azimuth between $\ptvecmiss$ and $\ptvecl$. The prompt backgrounds from triboson and associated Higgs boson production are estimated from simulation using the calculated cross sections at NLO and are henceforth referred to as the VVV and the $\PH+\cPX$ backgrounds, respectively. Similarly, the background from $\ttbar \PW$ and $\ttbar$\PZ is estimated from simulation and is referred to as the $\ttbar\cPV$ background.

\begin{figure*}[!ht]
  \centering
  \includegraphics[width=0.49\textwidth]{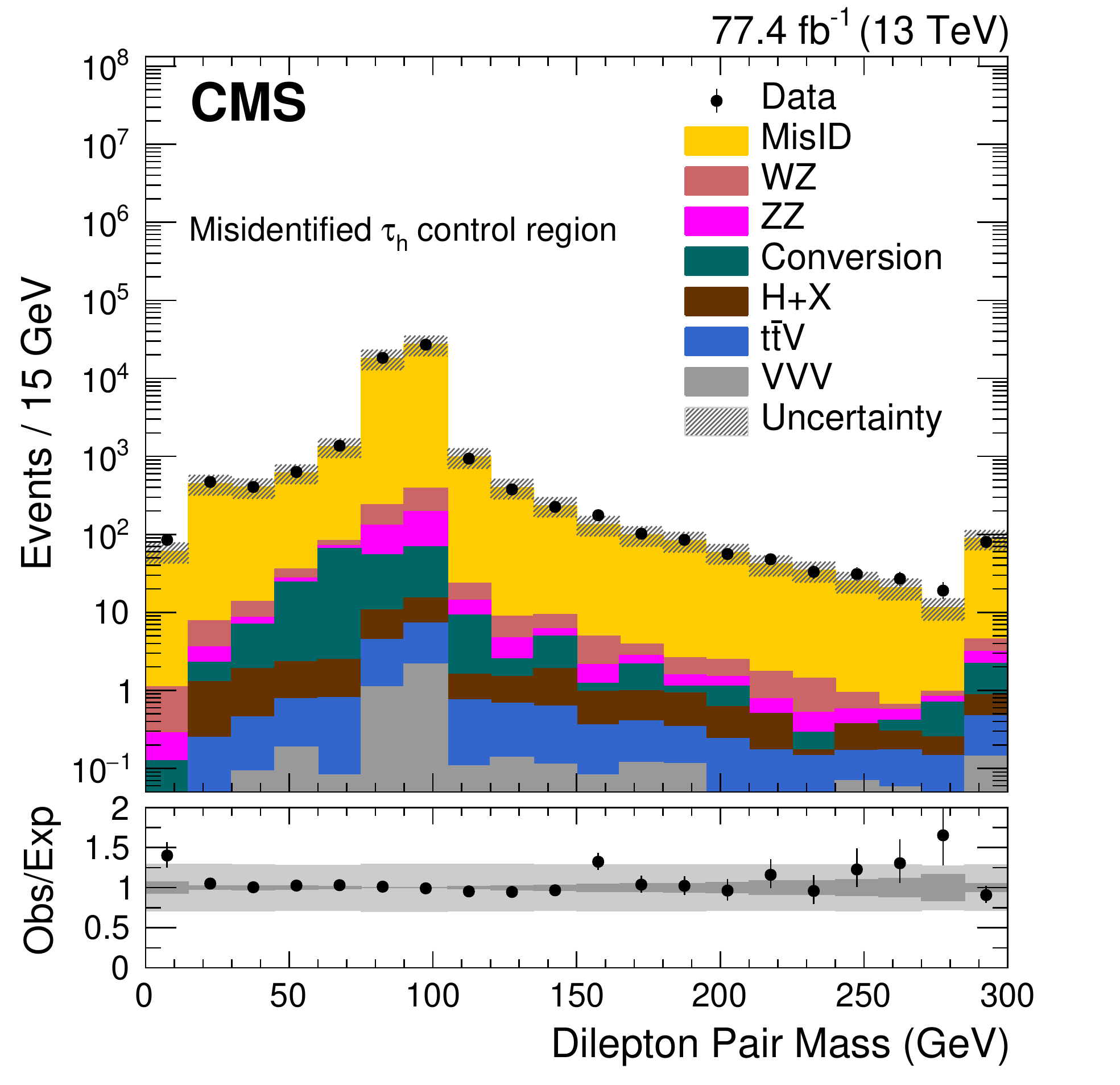}
  \includegraphics[width=0.49\textwidth]{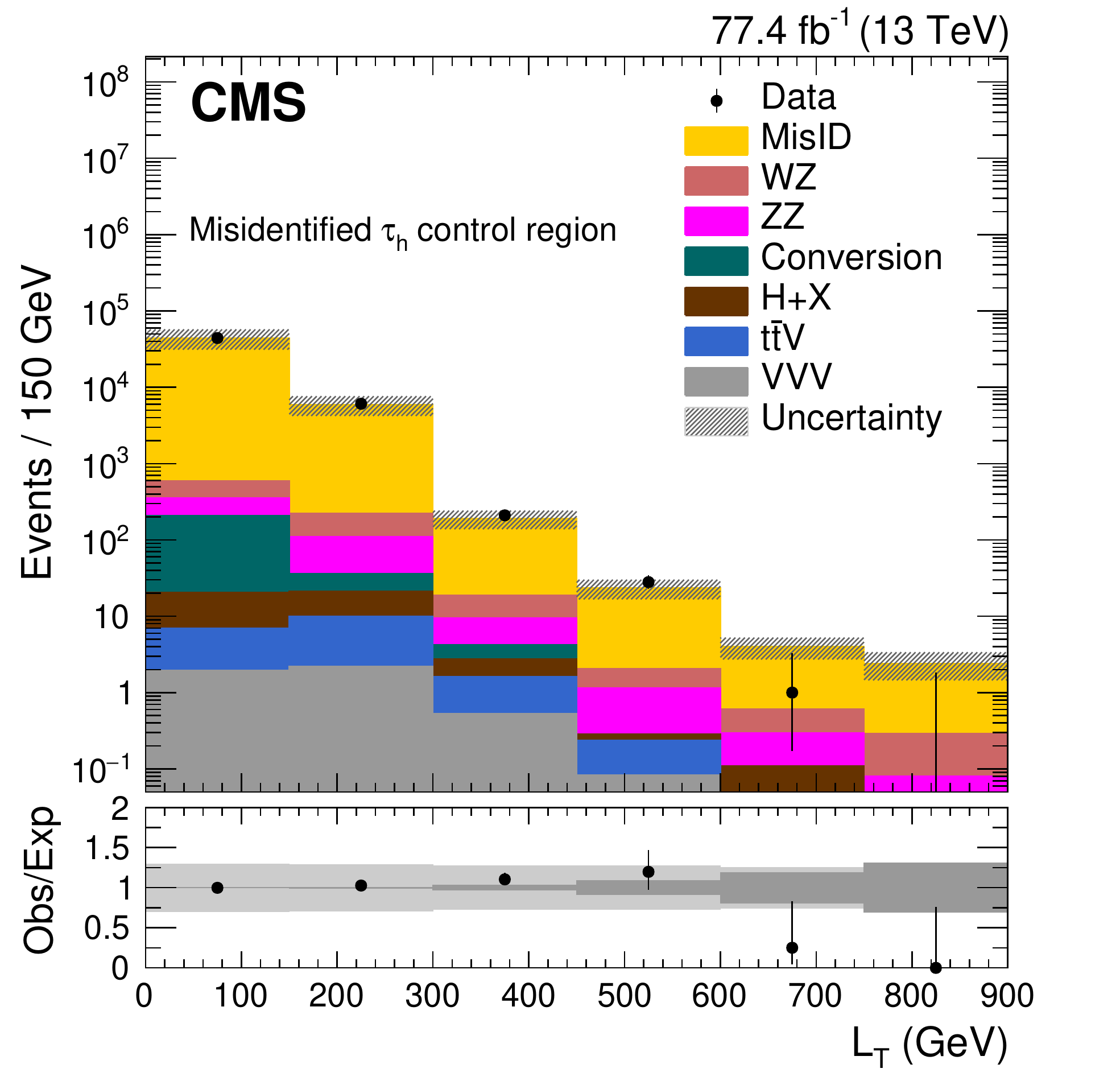}
\caption{\label{fig:fakes} The dilepton mass (left) and the \LT (right) distributions in data and simulation in a misidentified \tauh control region. This control region contains 2L1T (OS) events with $\ptmiss <50 \GeV$. The total SM background is shown as a stack of all contributing processes. The hatched gray bands in the upper panels represent the total uncertainty in the expected background. The lower panels show the ratios of observed data to the total expected background. In the lower panels, the light gray band represents the combined statistical and systematic uncertainty in the expected background, while the dark gray band represents the statistical uncertainty only. The rightmost bins include the overflow events.}
\end{figure*}

The MisID background arises from processes such as $\PZ +$jets and $\ttbar +$jets. This background is estimated using a 3-dimensional
implementation of a matrix method~\cite{Chatrchyan:2012bra}. In this method, rates are measured in data CRs for leptons to pass the analysis lepton selections, given that these leptons pass looser offline selections. It is assumed that these rates for prompt and misidentified leptons behave similarly
across the different CRs and SRs. We measure these rates in dedicated CRs: one with a dilepton
selection for prompt rates, and another with a trilepton signal-depleted selection with one OSSF on-\PZ pair and $\ptmiss < 50 \GeV$ for
misidentification rates. The rates are parameterized as functions of lepton $\pt$ and $\eta$. An additional correction factor is applied
as a function of the number of charged particles, to account for rate variations due to the hadronic activity in the event. For \tauh
misidentification rates, an additional parameterization is needed, based on the $\pt$ of the jet matched to the \tauh. This is required
to account correctly for rate variations due to the boost of the lepton system. The rate measurements are dominated by $\PZ+$jets events
and are corrected using simulation to an average of the $\PZ+$jets and $\ttbar +$jets events. Figure~\ref{fig:fakes} demonstrates the
agreement between the expected background and the observed data yields, as a function of the dilepton mass and \LT, in a signal-depleted 2L1T (OS) selection.

\section{Systematic uncertainties}\label{sec:systematics}

The primary sources of systematic uncertainty in the SM background arise from those in the MisID background, and from those in the \WZ and \ZZ backgrounds. The systematic uncertainty in the MisID background contribution arises primarily via the uncertainties in the measurement of prompt and misidentified rates in the matrix method. In addition, the uncertainties in the $\PZ+$jets and $\ttbar+$jets rates contribute to the systematic uncertainty in this background. We vary the rates within their respective uncertainties and observe the change in the background yield in all SRs. The final estimates vary by 20--35\% depending upon the year the data were collected and the SR. The \WZ and \ZZ background estimates have systematic uncertainties of 4--5\% arising from the normalization factor measurements in the dedicated CRs. The conversion background estimate has a systematic uncertainty of 11\%.

To account for differences between the data and simulation, a number of different sources of systematic uncertainty are considered. Lepton energy (or momentum) scale uncertainties, as well as jet and lepton resolution uncertainties are applied at the per-object level, where the corresponding object momenta are varied up and down by their corresponding uncertainties. This results in a 2--10\% impact on the background prediction, depending on \LT and the SR. The uncertainty in the trigger efficiency results in a 2--3\% uncertainty in the background prediction. Additionally, an integrated luminosity measurement uncertainty of 2.5 (2.3)\% is applied to the simulated rare background estimates for the 2016~\cite{CMS-PAS-LUM-17-001} (2017~\cite{CMS-PAS-LUM-17-004}) analysis. For the sub-dominant, rare background processes such as $\ttbar\cPV$, triboson, or associated Higgs boson production, a 50\% systematic uncertainty is applied to the theoretical cross sections to cover the PDF, and the renormalization and factorization scale uncertainties. The pileup modeling uncertainty is evaluated by varying the cross section used in the reweighting procedure up and down by 5\%, which results in a 4\% impact on background yields according to simulation. The typical variations for various sources of systematic uncertainty are provided in Table~\ref{tab:Systematics}.

\begin{table*}[!ht]
\centering
\topcaption{The sources of systematic uncertainty and the typical variations (\%) observed in the affected background and signal yields in the analysis. All sources of uncertainty are considered as correlated between the 2016 and 2017 data analyses except for the lepton identification and isolation, the single lepton trigger, and the integrated luminosity. The label ALL is defined as \WZ, \ZZ, Rare ($\ttbar\cPV$, VVV, Higgs boson), and Signal processes.}\label{tab:Systematics}
\begin{scotch}{lcc}
Source of uncertainty& Typical variations (\%)& Processes\\
\hline
MisID background            & 20--35        & \NA  \\
Rare background normalization                 & 50     & \NA   \\
Conversion background normalization             & 11     & \NA   \\
\WZ background normalization                                   & 5     & \NA   \\
\ZZ background normalization                                   & 4--5         & \NA   \\
Lepton identification \& isolation                            & 6--8 & ALL\\
Single lepton trigger                               & $<$3    & ALL\\
Electron energy scale and resolution                & 2--5    & ALL   \\
Muon momentum scale and resolution                 & 2--10   & ALL   \\
Hadronic \Pgt lepton energy scale                                    & $<$5    & ALL   \\
Jet energy scale                                    & 5--10     & ALL   \\
Unclustered energy scale                            & 1--10     & ALL   \\
Integrated luminosity                               & 2.3--2.5   & Rare/Signal \\
Pileup modeling                                              & $<$4   & ALL   \\
\end{scotch}
\end{table*}

\begin{figure*}[!ht]
  \centering

  \includegraphics[width=0.49\textwidth]{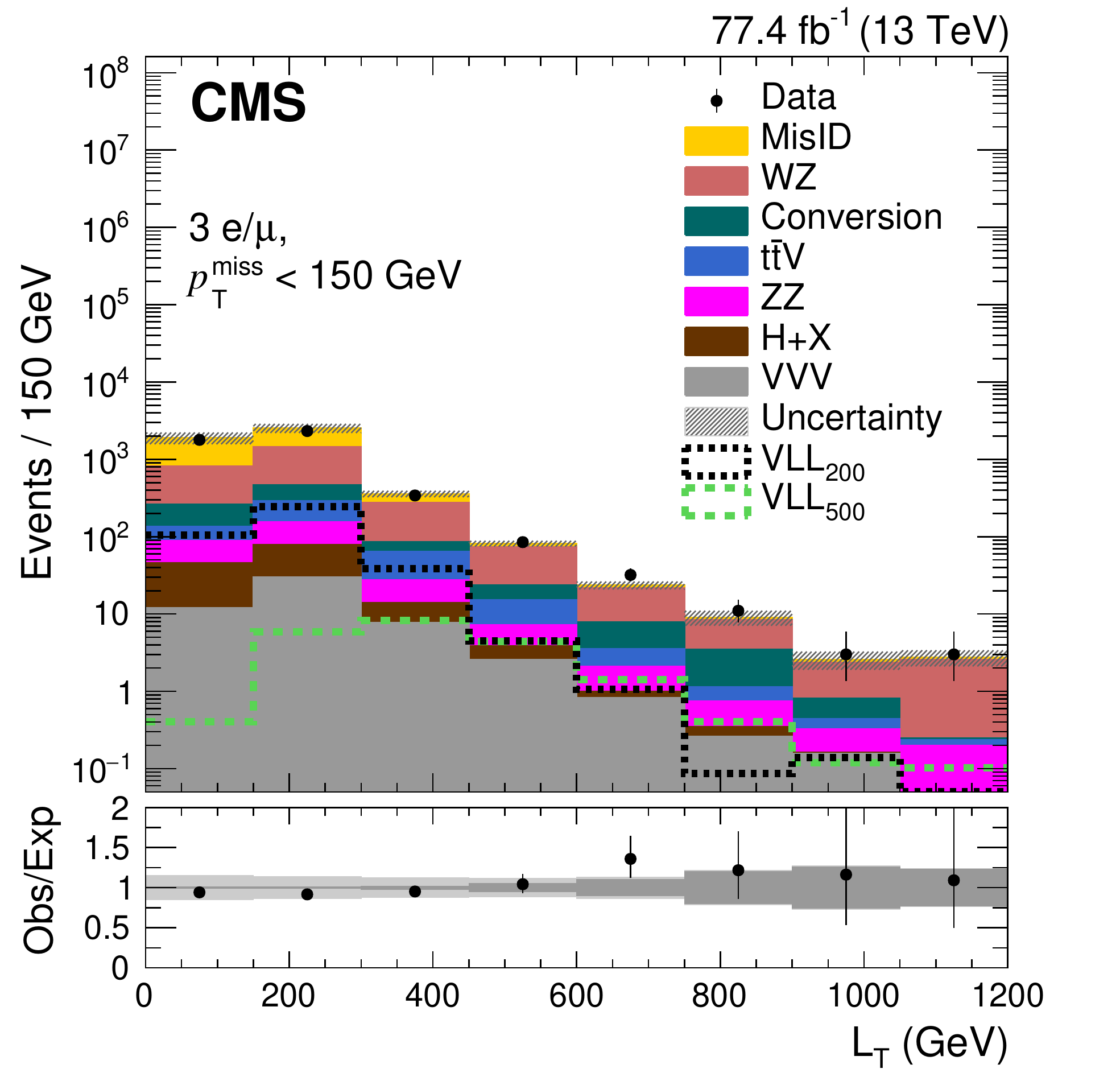}
  \includegraphics[width=0.49\textwidth]{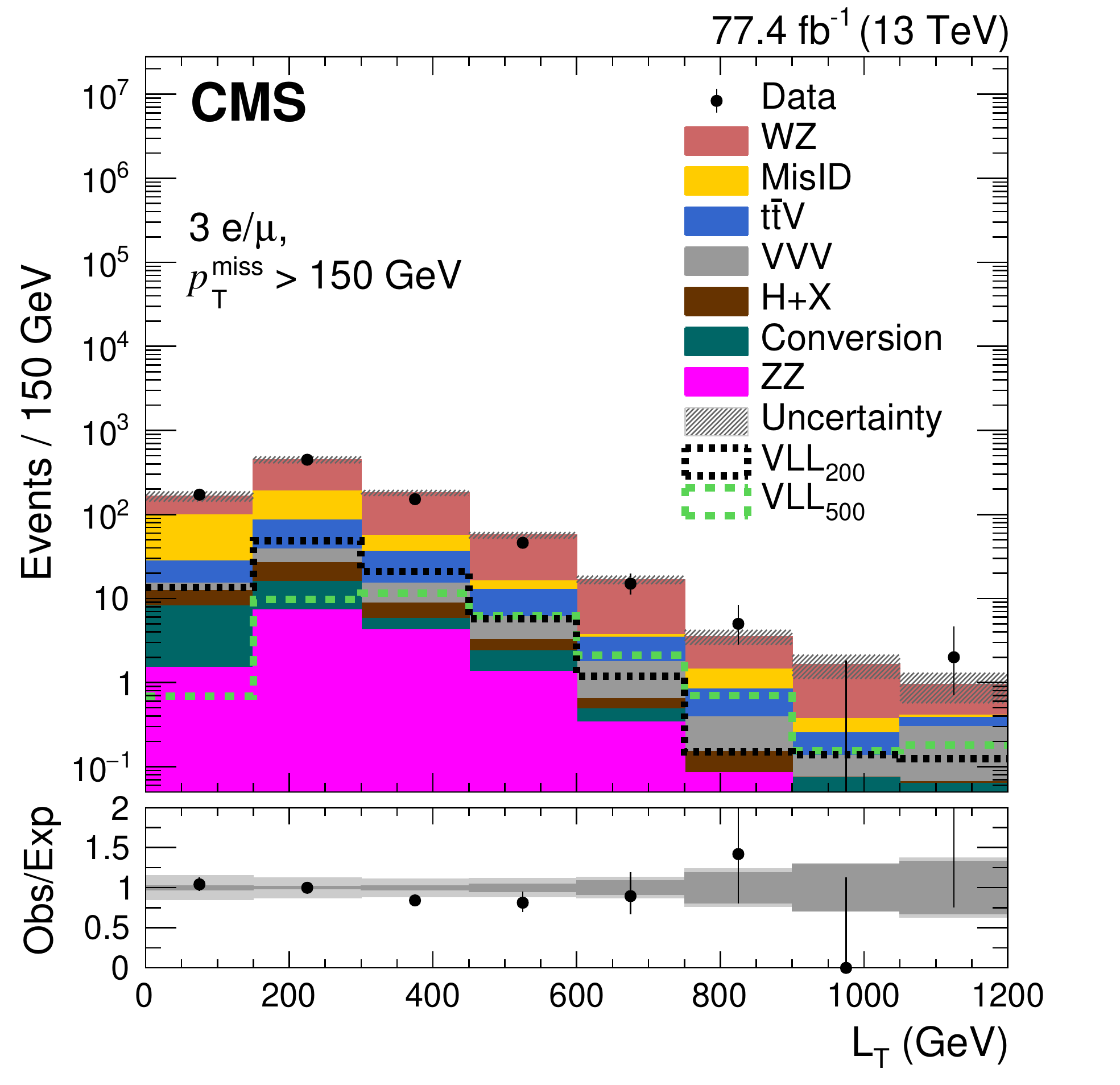}
  \includegraphics[width=0.49\textwidth]{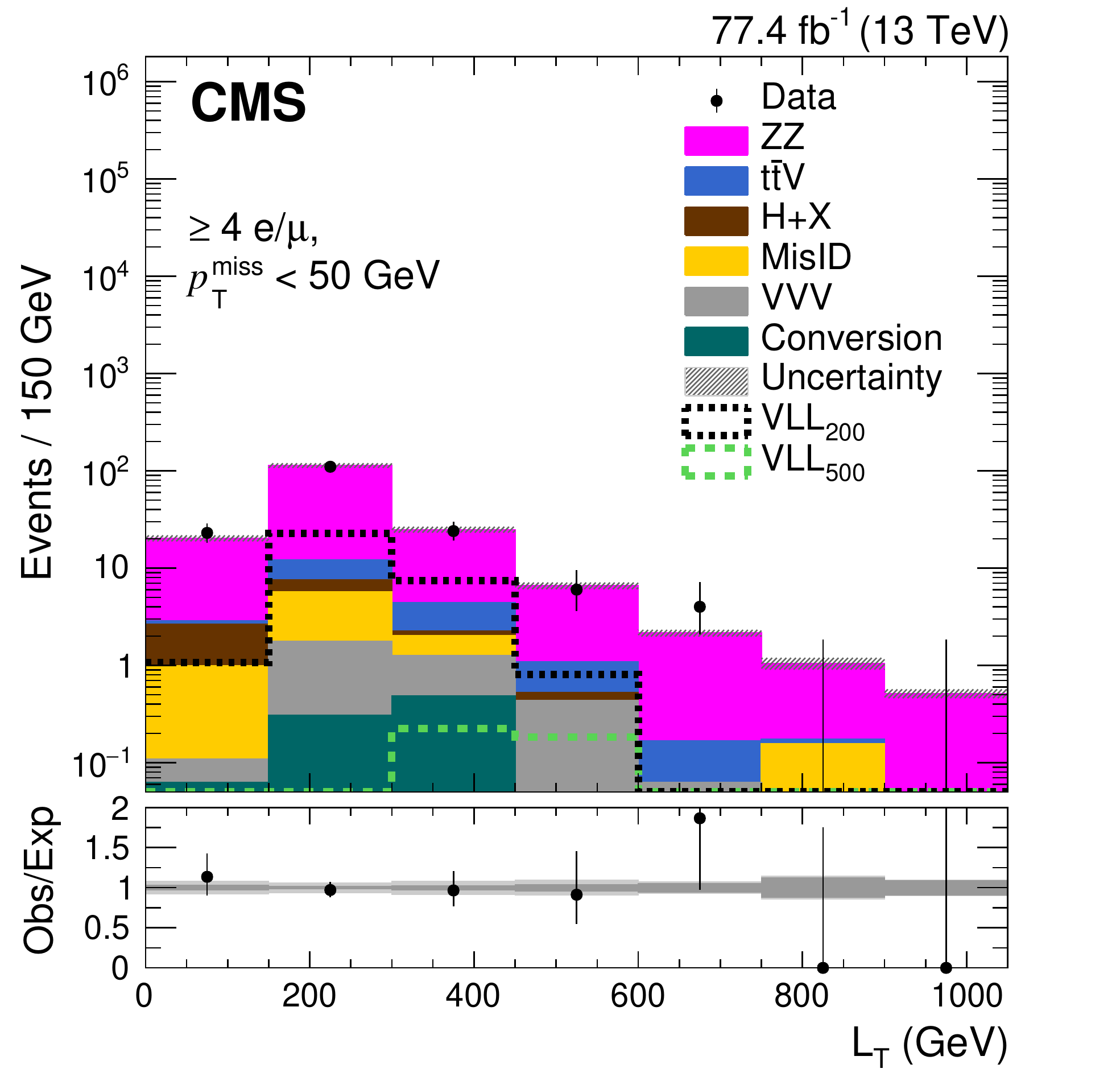}
  \includegraphics[width=0.49\textwidth]{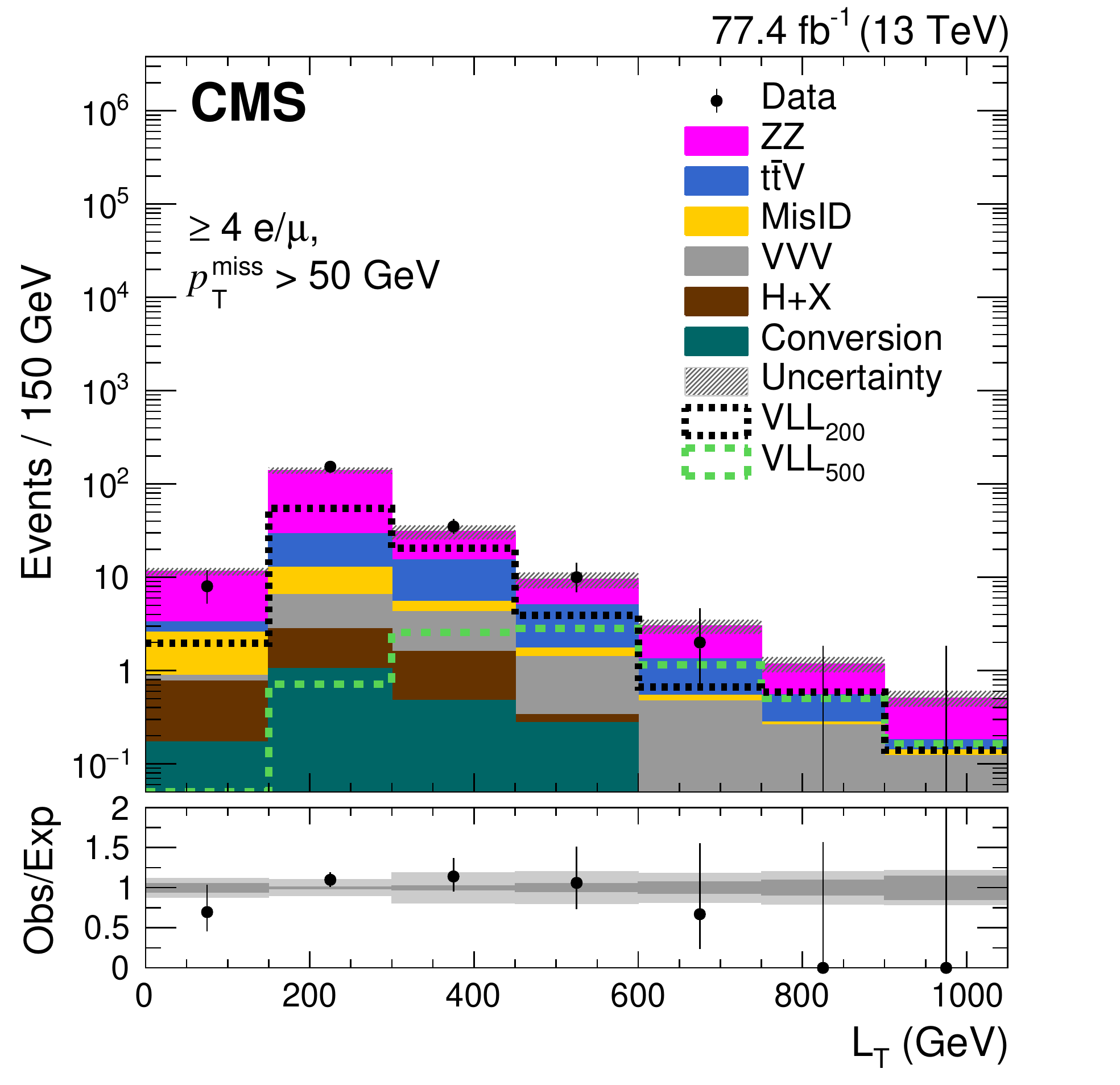}

  \caption{ The \LT distributions for the 3L signal regions with $\ptmiss<150 \GeV$ (upper left) and $\ptmiss>150 \GeV$ (upper right), and for the 4L signal regions with $\ptmiss<50 \GeV$ (lower left) and $\ptmiss>50 \GeV$ (lower right). The total SM background is shown as a stack of all contributing processes. The predictions for VLL signal models (the sum of all production and decay modes) with $m_{\Pgt'/\PGn'} = 200$ and 500\GeV are shown as dashed lines. The hatched gray bands in the upper panels represent the total uncertainty in the expected background. The lower panels show the ratios of observed data to the total expected background. In the lower panels, the light gray band represents the combined statistical and systematic uncertainty in the expected background, while the dark gray band represents the statistical uncertainty only. The rightmost bins include the overflow events.}\label{fig:3L4LSRF}
\end{figure*}

\begin{figure*}[!ht]
  \centering
  \includegraphics[width=0.49\textwidth]{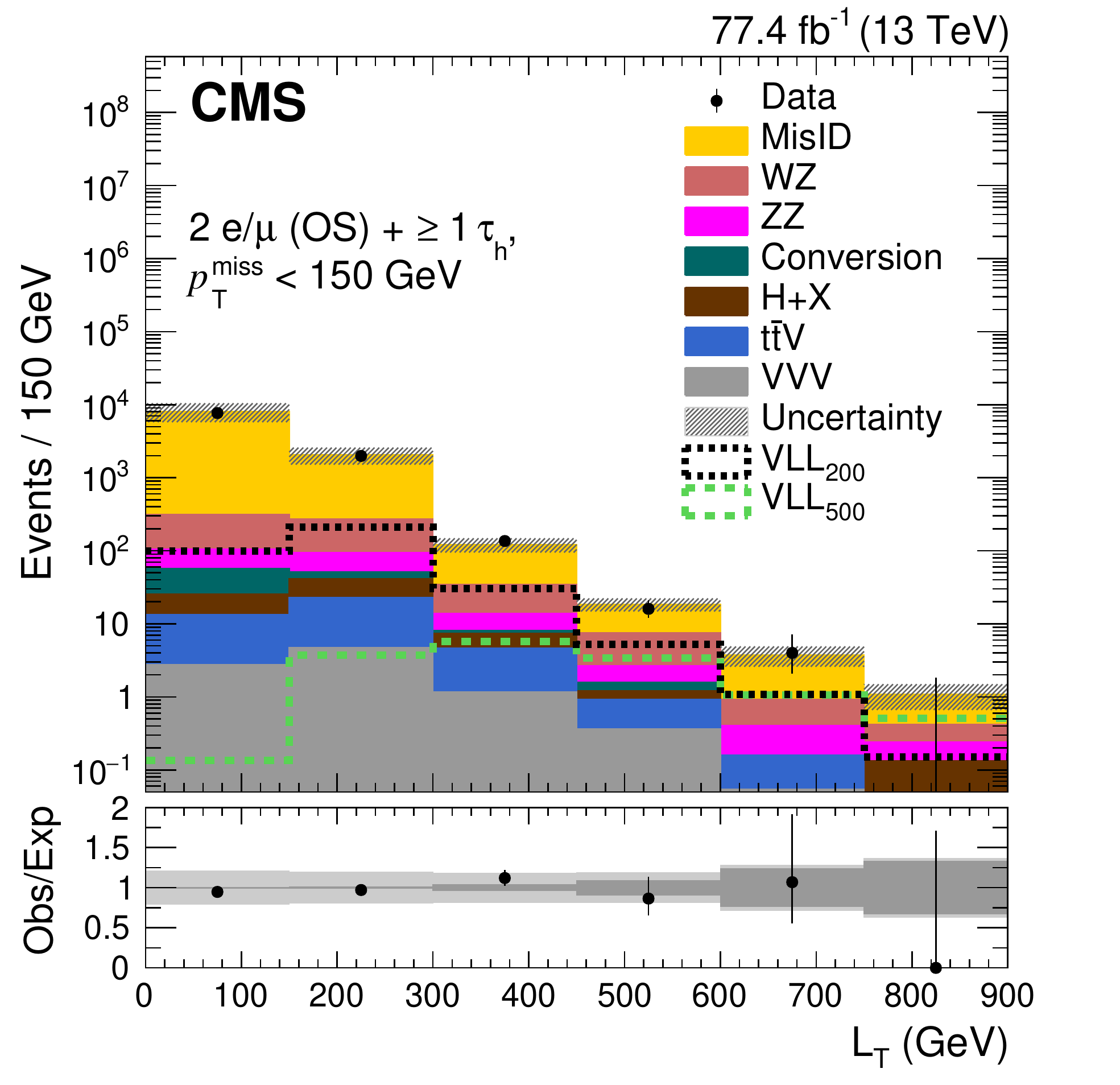}
  \includegraphics[width=0.49\textwidth]{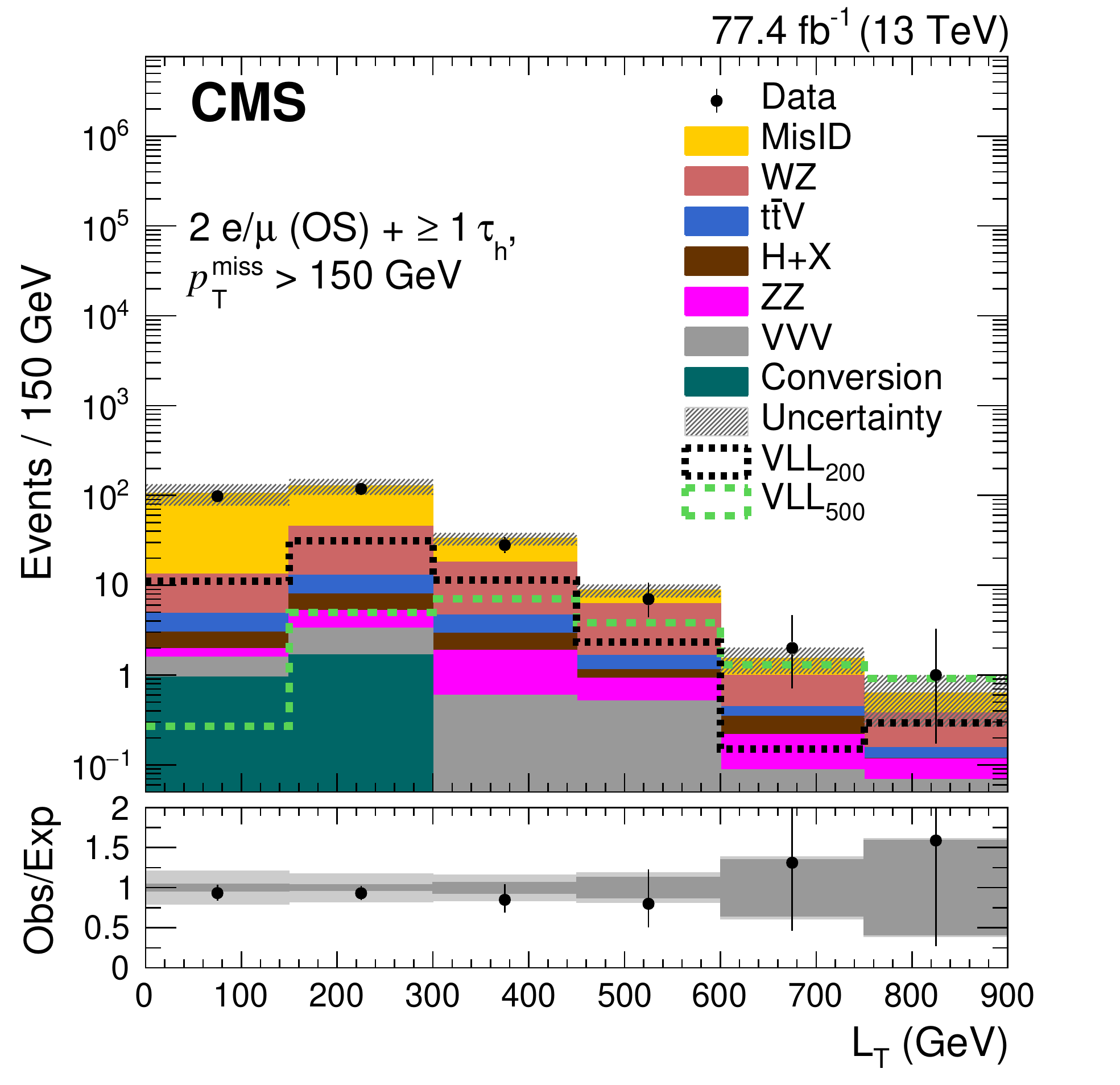}
  \includegraphics[width=0.49\textwidth]{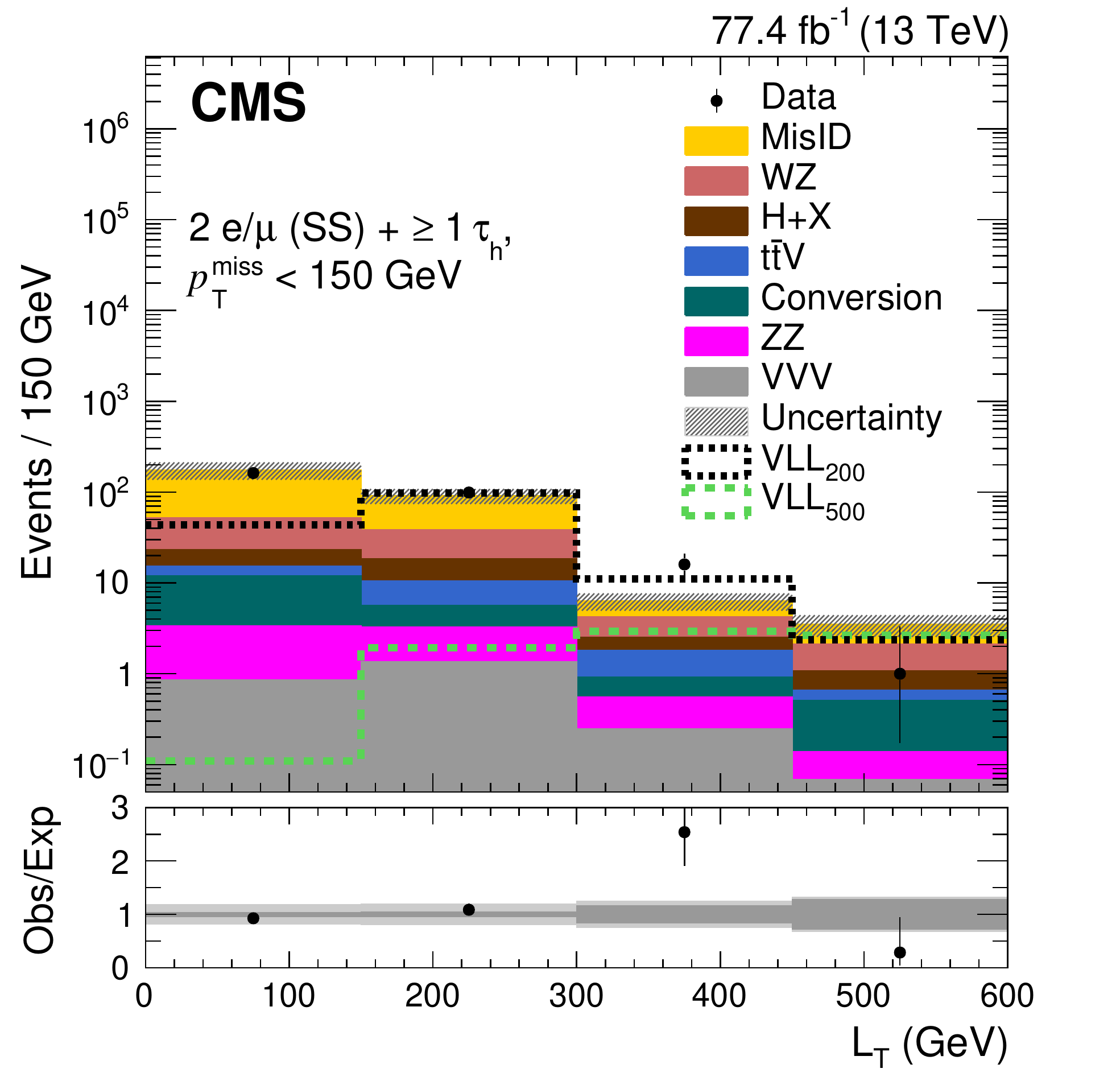}
  \includegraphics[width=0.49\textwidth]{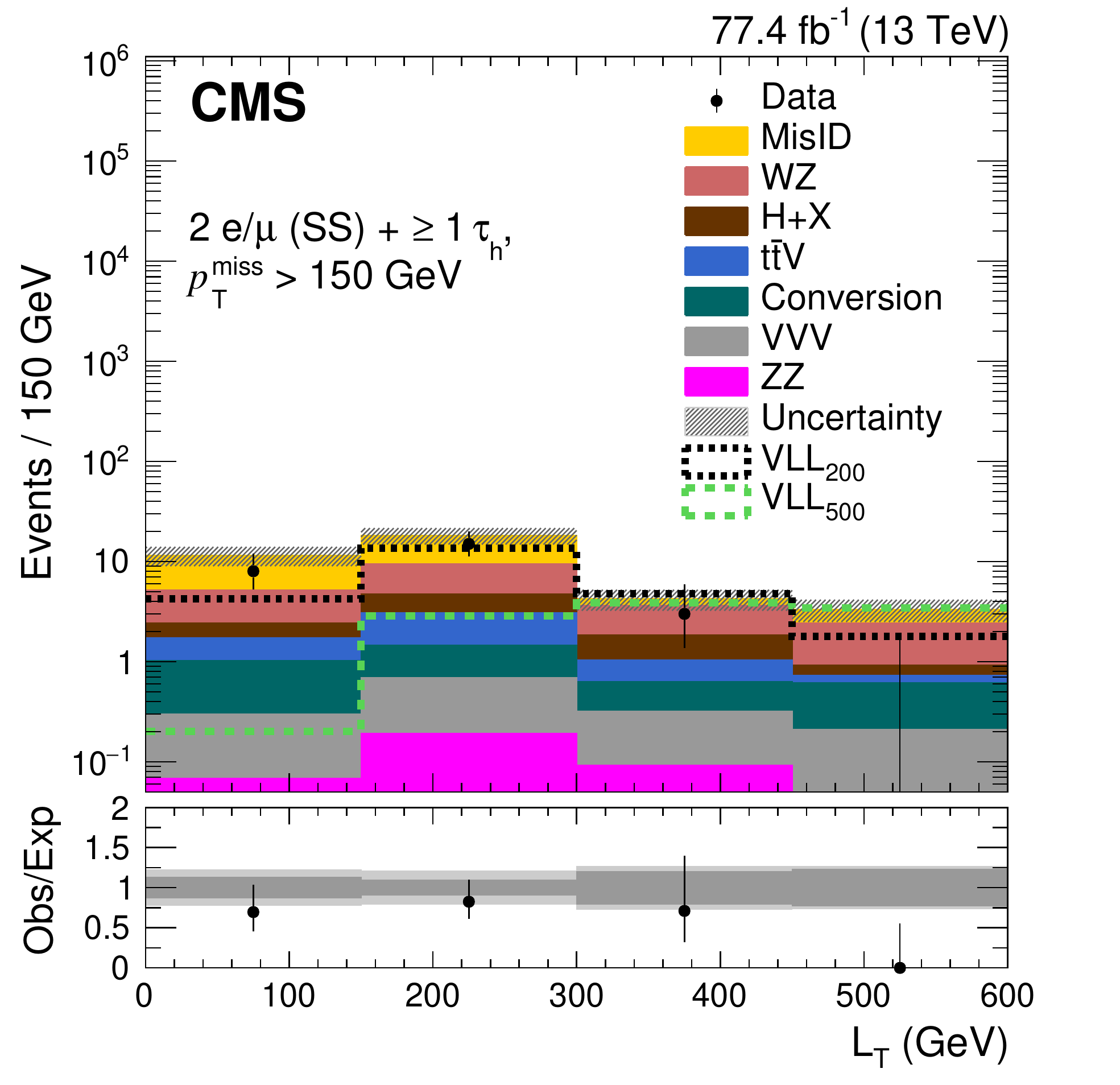}
    \caption{ The \LT distributions for the  2L1T OS signal regions with $\ptmiss<150 \GeV$ (upper left) and $\ptmiss>150 \GeV$ (upper right), and for the 2L1T SS signal
      regions with $\ptmiss<150 \GeV$ (lower left) and $\ptmiss>150 \GeV$ (lower right). The total SM background is shown as a stack of all contributing processes. The predictions for VLL signal models (sum of all production and decay modes) with $m_{\Pgt'/\PGn'} = 200$ and 500\GeV are also shown as dashed lines. The hatched gray bands in the upper panels represent the total uncertainty in the expected background. The lower panels show the ratios of observed data to the total expected background. In the lower panels, the light gray band represents the combined statistical and systematic uncertainty in the expected background, while the dark gray band represents the statistical uncertainty only. The rightmost bins include the overflow events.}\label{fig:OSSSRF}
\end{figure*}

\begin{figure}[!ht]
  \centering

  \includegraphics[width=0.49\textwidth]{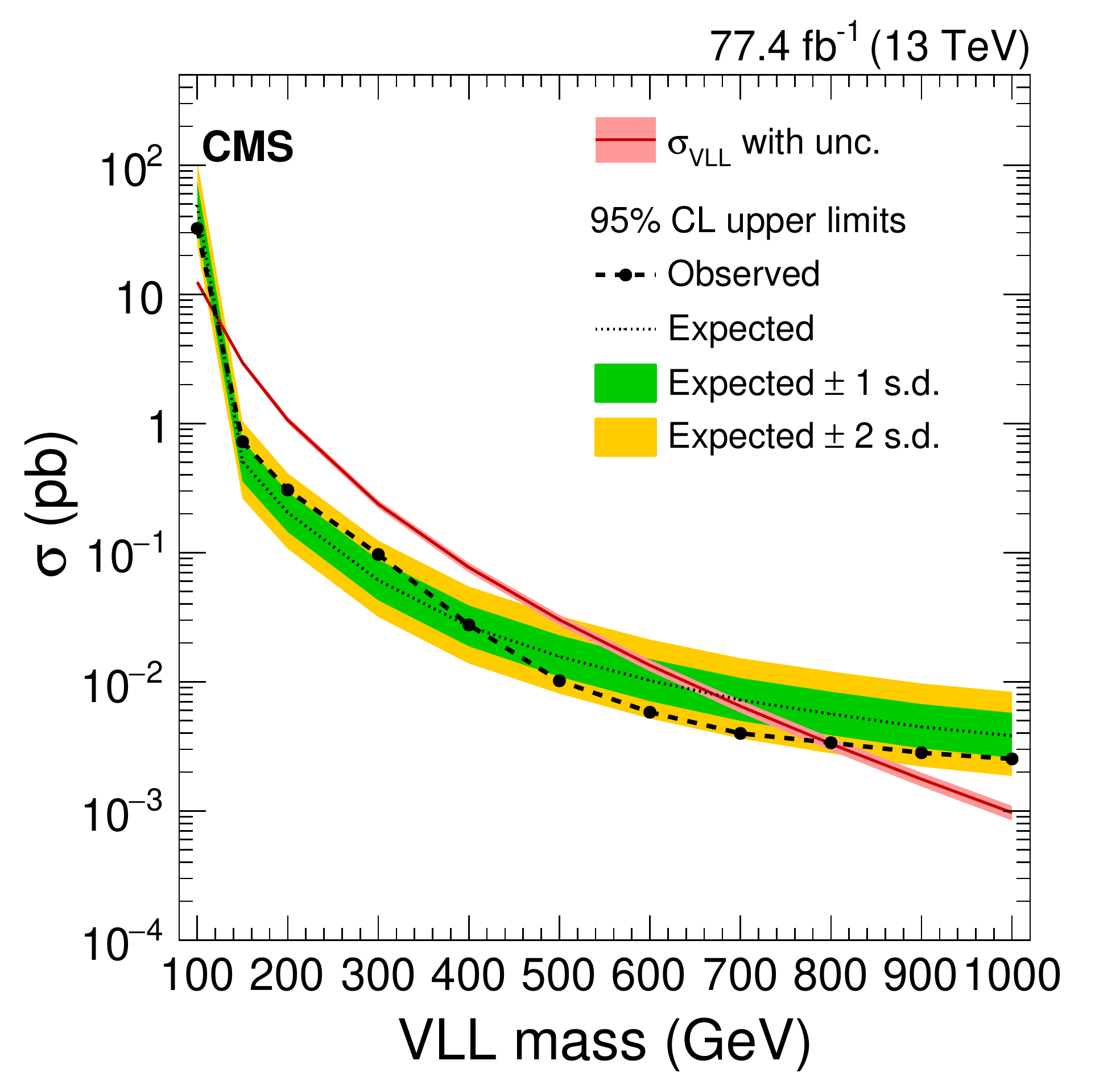}
  \caption{The 95$\%$ confidence level upper limits on the total cross section for associated ($\Pgt '^\pm \PGn^\prime_{\Pgt}$) and pair
    ($\Pgt '^{+} \Pgt '^{-}$/$\PGn^\prime_{\Pgt}\PAGn^\prime_{\Pgt}$)
    production of VLLs. Also shown is the theoretical prediction for the production cross section of a vector-like lepton doublet coupling to the third generation SM leptons. The observed (expected) exclusion limit on the masses of VLLs is in the range of 120--790 (120--680)\GeV.}
  \label{fig:limitsFull}
\end{figure}

\section{Results}\label{sec:conclusion}
The \LT distributions for the 4L and 3L SRs are shown in Fig.~\ref{fig:3L4LSRF}, while those for various 2L1T SRs are shown in Fig.~\ref{fig:OSSSRF}. We do not observe any significant discrepancies between the background predictions and the observed data. Limits are set on the combined cross section for associated ($\Pgt ' \PGn^\prime_{\Pgt}$) and pair ($\Pgt ' \Pgt '$/$\PGn^\prime_{\Pgt} \PGn^\prime_{\Pgt}$) production of VLLs. To obtain upper limits on the signal cross section at 95\% confidence level (\CL), we use a modified frequentist approach with a test statistic based on the profile likelihood in the asymptotic approximation and the \CLs criterion~\cite{JUNK1999435,cls,Cowan2011}. The upper limits are shown in Fig.~\ref{fig:limitsFull}. We use a linear interpolation of the expected event yields between the simulated signal samples in the limit calculations. Systematic uncertainties are incorporated into the likelihood as nuisance parameters with log-normal probability distributions, while statistical uncertainties are modeled with gamma functions. The observed limits are within two standard deviations of the expected limits from the background-only hypothesis. Because of the preferential coupling of VLLs to \Pgt leptons, the major contribution to these results comes from the 2L1T SRs. The analysis sensitivity benefits from the large signal-to-background ratio in the 2L1T (SS) SRs, despite the small production rate for this channel. The measurements in the 2L1T channels alone exclude VLLs in the mass range 120--740\GeV. On combining all the 4L, 3L, and 2L1T SRs, with the hypothesis of an SU(2) mass degenerate VLL doublet with couplings to the third generation SM leptons, we exclude VLLs with mass in the range of 120--790\GeV at 95\% $\CL$.

\section{Summary}\label{sec:summary}

A search for vector-like leptons coupled to the third generation standard model leptons has been performed in several multilepton final states using \fullLumi of proton-proton collision data at a center-of-mass energy of 13\TeV, collected by the CMS experiment in 2016 and 2017. No significant deviations of the data from the standard model predictions are observed. These results exclude a vector-like lepton doublet with a common mass in the range 120--790\GeV at 95\% confidence level. These are the most stringent limits yet on the production of a vector-like lepton doublet, coupling to the third generation standard model leptons.

\begin{acknowledgments}
We congratulate our colleagues in the CERN accelerator departments for the excellent performance of the LHC and thank the technical and administrative staffs at CERN and at other CMS institutes for their contributions to the success of the CMS effort. In addition, we gratefully acknowledge the computing centers and personnel of the Worldwide LHC Computing Grid for delivering so effectively the computing infrastructure essential to our analyses. Finally, we acknowledge the enduring support for the construction and operation of the LHC and the CMS detector provided by the following funding agencies: BMBWF and FWF (Austria); FNRS and FWO (Belgium); CNPq, CAPES, FAPERJ, FAPERGS, and FAPESP (Brazil); MES (Bulgaria); CERN; CAS, MoST, and NSFC (China); COLCIENCIAS (Colombia); MSES and CSF (Croatia); RPF (Cyprus); SENESCYT (Ecuador); MoER, ERC IUT, PUT and ERDF (Estonia); Academy of Finland, MEC, and HIP (Finland); CEA and CNRS/IN2P3 (France); BMBF, DFG, and HGF (Germany); GSRT (Greece); NKFIA (Hungary); DAE and DST (India); IPM (Iran); SFI (Ireland); INFN (Italy); MSIP and NRF (Republic of Korea); MES (Latvia); LAS (Lithuania); MOE and UM (Malaysia); BUAP, CINVESTAV, CONACYT, LNS, SEP, and UASLP-FAI (Mexico); MOS (Montenegro); MBIE (New Zealand); PAEC (Pakistan); MSHE and NSC (Poland); FCT (Portugal); JINR (Dubna); MON, RosAtom, RAS, RFBR, and NRC KI (Russia); MESTD (Serbia); SEIDI, CPAN, PCTI, and FEDER (Spain); MOSTR (Sri Lanka); Swiss Funding Agencies (Switzerland); MST (Taipei); ThEPCenter, IPST, STAR, and NSTDA (Thailand); TUBITAK and TAEK (Turkey); NASU and SFFR (Ukraine); STFC (United Kingdom); DOE and NSF (USA).

\hyphenation{Rachada-pisek} Individuals have received support from the Marie-Curie program and the European Research Council and Horizon 2020 Grant, contract Nos.\ 675440, 752730, and 765710 (European Union); the Leventis Foundation; the A.P.\ Sloan Foundation; the Alexander von Humboldt Foundation; the Belgian Federal Science Policy Office; the Fonds pour la Formation \`a la Recherche dans l'Industrie et dans l'Agriculture (FRIA-Belgium); the Agentschap voor Innovatie door Wetenschap en Technologie (IWT-Belgium); the F.R.S.-FNRS and FWO (Belgium) under the ``Excellence of Science -- EOS" -- be.h project n.\ 30820817; the Beijing Municipal Science \& Technology Commission, No. Z181100004218003; the Ministry of Education, Youth and Sports (MEYS) of the Czech Republic; the Lend\"ulet (``Momentum") Program and the J\'anos Bolyai Research Scholarship of the Hungarian Academy of Sciences, the New National Excellence Program \'UNKP, the NKFIA research grants 123842, 123959, 124845, 124850, 125105, 128713, 128786, and 129058 (Hungary); the Council of Science and Industrial Research, India; the HOMING PLUS program of the Foundation for Polish Science, cofinanced from European Union, Regional Development Fund, the Mobility Plus program of the Ministry of Science and Higher Education, the National Science Center (Poland), contracts Harmonia 2014/14/M/ST2/00428, Opus 2014/13/B/ST2/02543, 2014/15/B/ST2/03998, and 2015/19/B/ST2/02861, Sonata-bis 2012/07/E/ST2/01406; the National Priorities Research Program by Qatar National Research Fund; the Programa Estatal de Fomento de la Investigaci{\'o}n Cient{\'i}fica y T{\'e}cnica de Excelencia Mar\'{\i}a de Maeztu, grant MDM-2015-0509 and the Programa Severo Ochoa del Principado de Asturias; the Thalis and Aristeia programs cofinanced by EU-ESF and the Greek NSRF; the Rachadapisek Sompot Fund for Postdoctoral Fellowship, Chulalongkorn University and the Chulalongkorn Academic into Its 2nd Century Project Advancement Project (Thailand); the Welch Foundation, contract C-1845; and the Weston Havens Foundation (USA).
\end{acknowledgments}

\bibliography{auto_generated}  
\cleardoublepage \appendix\section{The CMS Collaboration \label{app:collab}}\begin{sloppypar}\hyphenpenalty=5000\widowpenalty=500\clubpenalty=5000\vskip\cmsinstskip
\textbf{Yerevan Physics Institute, Yerevan, Armenia}\\*[0pt]
A.M.~Sirunyan$^{\textrm{\dag}}$, A.~Tumasyan
\vskip\cmsinstskip
\textbf{Institut für Hochenergiephysik, Wien, Austria}\\*[0pt]
W.~Adam, F.~Ambrogi, T.~Bergauer, J.~Brandstetter, M.~Dragicevic, J.~Erö, A.~Escalante~Del~Valle, M.~Flechl, R.~Frühwirth\cmsAuthorMark{1}, M.~Jeitler\cmsAuthorMark{1}, N.~Krammer, I.~Krätschmer, D.~Liko, T.~Madlener, I.~Mikulec, N.~Rad, J.~Schieck\cmsAuthorMark{1}, R.~Schöfbeck, M.~Spanring, D.~Spitzbart, W.~Waltenberger, C.-E.~Wulz\cmsAuthorMark{1}, M.~Zarucki
% ...
\vskip\cmsinstskip
\dag: Deceased\\
1:  Also at Vienna University of Technology, Vienna, Austria\\
% ...
\end{sloppypar}
\end{document}